\documentclass[journal]{IEEEtran}

\usepackage{amsmath,graphicx}
\usepackage{booktabs}
\usepackage{multirow}
\usepackage{multicol}
\usepackage{colortbl}
\usepackage{subfigure}
\usepackage{cleveref}
\usepackage{enumerate}
\usepackage{fancybox}
\usepackage{cleveref}

\begin{document}
%

\title{Intermediate Deep Feature Compression: the Next Battlefield of Intelligent Sensing}
%
%
%

\author{Zhuo~Chen$^{1}$,
        Weisi~Lin$^{2}$,~\IEEEmembership{Fellow,~IEEE,}
        Shiqi~Wang$^{3}$,
        Lingyu~Duan$^{4}$,
        and~Alex~C.~Kot$^{5}$,~\IEEEmembership{Fellow,~IEEE}

{\small\begin{minipage}{\linewidth}\begin{center}
\begin{tabular}{ccc}
$^{1}$Interdisciplinary Graduate School, Nanyang Technological University \\
$^{2}$School of Computer Science and Engineering, Nanyang Technological University \\
$^{3}$Department of Computer Science, City University of Hong Kong \\
$^{4}$Institute of Digital Media, Peking University, Beijing, China \\
$^{5}$School of Electrical and Electronic Engineering, Nanyang Technological University \\
\end{tabular}
\end{center}\end{minipage}}
}


\maketitle

\begin{abstract}
The recent advances of hardware technology have made the intelligent analysis equipped at the front-end with deep learning more prevailing and practical. To better enable the intelligent sensing at the front-end, instead of compressing and transmitting visual signals or the ultimately utilized top-layer deep learning features, we propose to compactly represent and convey the intermediate-layer deep learning features of high generalization capability, to facilitate the collaborating approach between front and cloud ends. This strategy enables a good balance among the computational load, transmission load and the generalization ability for cloud servers when deploying the deep neural networks for large scale cloud based visual analysis. Moreover, the presented strategy also makes the standardization of deep feature coding more feasible and promising, as a series of tasks can simultaneously benefit from the transmitted intermediate layers. We also present the results for evaluation of lossless deep feature compression with four benchmark data compression methods, which provides meaningful investigations and baselines for future research and standardization activities.
\end{abstract}


%
\IEEEpeerreviewmaketitle

\section{Introduction}
\label{sec:intro}
\IEEEPARstart{R}{ecently}, deep neural networks (DNNs) have demonstrated the state-of-the-art performance in various computer vision tasks, e.g., image classification \cite{krizhevsky2012imagenet,simonyan2014very,he2016deep,huang2017densely}, image object detection \cite{girshick2016region,redmon2018yolov3}, visual tracking \cite{wang2015visual}, visual retrieval \cite{lin2017hnip}.
In contrast to the handcrafted features such as Scale-Invariant Feature Transform (SIFT) \cite{lowe2004distinctive}, deep learning based approaches are able to learn representative features directly from the vast amounts of data. 
For image classification, which is the fundamental task of computer vision, the AlexNet model \cite{krizhevsky2012imagenet} has achieved 9\% better classification accuracy than the previous hand-crafted methods in the 2012 ImageNet competition \cite{russakovsky2015imagenet}, which provides a large scale training dataset with 1.2 million images and one thousand categories. Inspired by the fantastic progress of AlexNet, DNN models continue to be the undisputed leaders in the competition of ImageNet. In particular, both VGGNet \cite{simonyan2014very} and GoogLeNet \cite{szegedy2015going} announced promising performance in the ILSVRC 2014 classification challenge, which demonstrated that deeper and wider architectures can bring great benefits in learning better representations via large scale datasets. In 2016, He \textit{et al.} also proposed residual blocks to enable very deep learning structure \cite{he2016deep}.

\begin{figure}[t]
\centering
\includegraphics[width=0.45\textwidth]{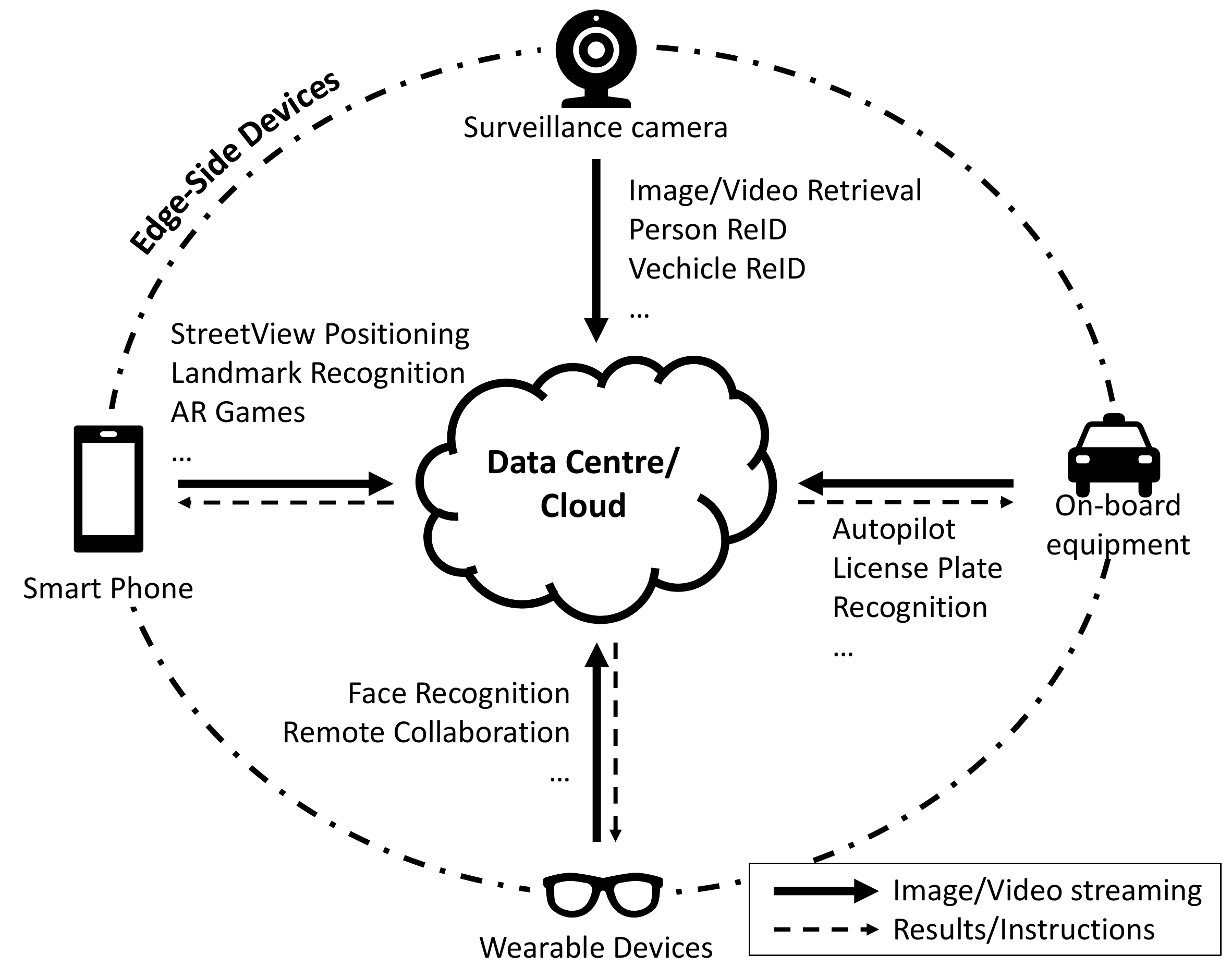}
\caption{Diagram of cloud-based visual analysis applications. Images and videos are acquired on the front-end and the analysis is performed at the cloud end. The two sides collaborate together through data transmission.}
\label{fig:cloudbase}
\end{figure}

With the advances of network infrastructure, cloud-based applications are springing up in recent years. In particular, the front-end devices acquire information from users or the physical world, which are subsequently transmitted to the cloud end (i.e., data center) for further process and analyses. In particular, for visual analysis, the  front-end devices deployed in the real world such as surveillance cameras and wearable devices acquire massive visual data which are transmitted to the cloud side for analyses, as shown in Fig.~\ref{fig:cloudbase}. Many computer vision models powered by deep learning can be applied in such cloud-based paradigm, such as pedestrian detection \cite{ouyang2013joint}, person \cite{xiao2016learning} and vehicle re-identification \cite{liu2016deep} in surveillance systems; autopilot \cite{bojarski2016end} and license plate recognition \cite{polishetty2016next} with on-board devices; face recognition \cite{sun2014deep,taigman2014deepface}, landmark retrieval\cite{wang2017effective} and object detection \cite{girshick2016region,redmon2018yolov3} in portable device (e.g., mobile, smart glasses) applications. 

For data communication between front-end and cloud sever, video compression and transmission serve as the foundation infrastructure in the traditional ``compress-then-analyse'' paradigm. In other words, the front-end devices capture and compress the visual data at the signal level, such that the coding bitstream can be transmitted to the cloud server for analyses. After the decoding process at the cloud side, the feature extraction and visual analysis are subsequently performed. However, the vast amount of front-end devices produce thousands-of-thousands bitstreams simultaneously, especially in the scenarios of video surveillance and Internet-of-Things (IoT). The signal level visual compression imposes high transmission burden, which is usually unfordable in practical applications. Moreover, the computational load of the numerous deep learning models executed simultaneously for feature extraction also becomes a significant bottleneck for scaling up at the cloud server. 

An alternative strategy ``analyse-then-compress'', the rational of which lies in compressing and transmitting the features extracted at the front-end to the cloud center, provides a feasible solution as features instead of the visual signals are ultimately used for analysis. For hand-crafted features, the standards from MPEG including MPEG-CDVS \cite{duan2016overview} and MPEG-CDVA \cite{duan2017compact} specify the feature extraction and compression processes. For deep learning features, top-layer features of the deep learning models are usually transmitted to the cloud side, since the top-layer features of deep models are compact and can be straightforwardly utilized for analyses. For instance, in the face recognition task, the deep feature of a human face is only with dimension of 4K in Facebook DeepFace \cite{taigman2014deepface}, 128 in Google FaceNet \cite{schroff2015facenet}, and 300 in SenseTime DeepID3 \cite{sun2015deepid3}. In such scenarios, only the lightweight operations such as feature comparison are required to be performed at the cloud servers, while the heavy workloads of feature extraction are distributed to the front-end. Moreover, transmitting features is also favorable for privacy protection. In particular, instead of directly conveying the visual signal which may easily expose privacy, feature communication can largely avoid the disclosing of the visible information.

However, one obstacle that potentially hinders the applications of deep learning feature compression is that deep learning models are normally designed and trained for specific tasks, and the top-layer features are extraordinary abstract and task-specific, making such compressed features difficult to generalize. 
This also prevents the applications of the future standardization of the deep feature coding, as the standardized compact deep features shall be well generalized to enable the interoperability in different application scenarios. In view of this, the intermediate layer feature compression, which shifts the computing load while maintaining the availability of various visual analysis tasks is presented in this paper. The presented approach can be regarded as a compromise between the two extremes ``analysis-then-compression'' and ``compression-then-analysis'', and provides a good balance among the computational load, communication cost and the generalization ability.

The rest of the paper is organized as follows. 
Section~\ref{sec:related} provides a brief review on the compact visual information representation, including video compression and feature compression. 
Section~\ref{sec:transmit_feature} describes our proposed collaborating approach for cloud-based visual analysis applications.
In Section~\ref{sec:feature_compression}, we discuss and envision the future standardization of deep feature coding.
Section~\ref{sec:evaluation} presents the evaluation results of deep feature compression.
Finally, Section~\ref{sec:conclusion} concludes this paper.

\section{Related works}
\label{sec:related}

\subsection{Video Coding Standard}
High Efficiency Video Coding (HEVC) \cite{sullivan2012overview} is the state-of-the-art video coding standard, which achieves 50\% bit-rate reductions for equal perceptual visual quality comparing to H.264/MPEG-4 Advanced Video Coding (AVC) \cite{wiegand2003overview}. As a joint video project of ITU-T Video Coding Experts Group (VCEG) and the ISO/IEC Moving Picture Experts Group (MPEG), the standardization of HEVC was finalized in Jan. 2013. As a video coding standard, HEVC only specifies the decoder. In other words, the decoder conforming to the standard can correctly reconstruct the video based on the bitstream, and the encoder can be feasibly optimized according to the application scenarios and requirements. HEVC can be applied to both image and video in lossy and lossless ways. Recently, in Apr.2018, the standardization for new generation video coding, Versatile Video Coding (VVC), was launched. It is expected to be completed before 2020, with much superior coding performance compared to HEVC.

\subsection{Standardization of Handcrafted Feature Coding}
To provide a standardized bitstream syntax to enable interoperability in the context of image retrieval applications, MPEG published Compact Descriptors for Visual Search (CDVS) \cite{duan2016overview} in Sep. 2015.
CDVS leverages handcrafted local (i.e.,SIFT descriptors) and global (i.e. Scalable Compressed Fisher Vector) features to represent the visual characteristics of images. 
To achieve compact image representation while maintaining the discrimination capability, a series of compression techniques were developed. In particular, with the process of local feature selection, descriptor compression, location compression and descriptor aggregation, CDVS supports interoperability between different bitstream sizes by setting six operating points from 512B to 16KB.

Based on CDVS, MPEG has moved forward to the standardization of Compact Descriptors for Video Analysis (CDVA) \cite{duan2017compact} since Feb. 2015. Considering the fact that extracting the feature from each frame of videos leads to extremely high computational costs and redundancy in the video representations, multi-keyframe based retrieval strategy was adopted by the ongoing CDVA standard. More specifically, to generate the compact video descriptors, the local and global descriptors of sampled keyframes of the given video are firstly extracted by standardized CDVS. Then, these frame-level features are further compressed and packed to constitute the CDVA descriptors. Moreover, the deep learning features were also adopted into the working draft of CDVA to further boost the retrieval performance \cite{lin2017hnip}.  

\subsection{Deep Neural Network Compression}
Deep neural networks are featured by memory and computational intensive operations, which limit their applications in embedded systems. To address this issue, deep neural network compression has been widely investigated. 
The compression methods can be divided into three categories: weight quantization, network pruning and knowledge distillation. 
More specifically, weight quantization aims to compress deep learning models at the level of individual neuron weights. As such, all the parameters of a deep model can be maintained, while each weight is represented with less bits. The state-of-art weight quantization approach can represent one parameter with only two or three bits \cite{li2016ternary,rastegari2016xnor}, while maintaining high classification performance.
Network pruning aims to shrink the model size by zeroing out selected neuron weights \cite{lecun1990optimal}. Furthermore, it is demonstrated that this approach not only provides a feasible way to compress deep neural networks, but also helps to deal with the overfitting problems \cite{hassibi1993optimal}. A recent work \cite{han2015learning} shows that pruning the weights can compress the deep models without any performance drop.
Another branch of the compression approaches is knowledge distillation, which attempts to train a small student network to replace the large teacher network. In \cite{buciluǎ2006model,ba2014deep}, the authors trained shallow student networks to regress the top layer outputs of teacher networks, while in \cite{romero2014fitnets} a thin one was trained. In \cite{hinton2015distilling}, the authors modified the last layer logits of the teacher network to provide the student network with more information regarding how the teacher model generalizes.
It is worth mentioning that all these three types of approaches can be applied together to achieve 35x-49x reduction in terms of the volume of the state-of-the-art deep learning models~\cite{han2015deep}.

\subsection{Deep Learning Feature Compression}
\subsubsection{Compact Deep Representations}
In computer vision, visual embeddings from deep neural networks have been wildly used. To achieve compact and discriminative representations, existing methods can be classified into two categories. The first one aims to design the deep models with small-size embedding layers before training, and the other targets to add a series of dimension-reduction/binarization layers (e.g., hashing and PCA) on top of the trained deep learning models.
For the first category, the work in \cite{schroff2015facenet} explored the effect of the embedding dimensionality in deep face recognition models, and better image retrieval performance with smaller embedding size by tailoring the CNN architecture was achieved in~\cite{gordo2016deep}. 
For the second category, the authors in \cite{babenko2014neural} applied PCA compression on the top layer representations of a pre-trained CNN to achieve state-of-the-art accuracy on a number of image retrieval datasets. Moreover, hashing also plays an important role in deep embedding compression, and different hashing methods on the top of deep neural networks have been investigated~\cite{xia2014supervised,lin2015deep,zhao2015deep,lai2015simultaneous}. 

\subsubsection{Compression for Deep Features}
The deep feature compression aims to compress the extracted features from an off-the-shelf deep neural network in a restorable way for further usage. It is worth noting that, such feature can be the activations of any layers of a deep model.
In recent works \cite{choi2018deep,choi2018near}, the deep feature compression was investigated in the context of collaborative intelligence and image object detection. In particular, the work in \cite{choi2018deep} employed HEVC Range extension (RExt) to compress deep features extracted by two specific layers (i.e. Max11 and Max17) of the YOLO9000 network. Subsequently, the authors in \cite{choi2018near} proposed a near-lossless deep feature compressor and evaluated the performance on four deep networks.


\section{Towards the transmission of deep learning features}
\label{sec:transmit_feature}

\begin{figure}[t]
\centering
\subfigure[Visual signal transmission. By transmitting the visual signal, a series of visual analysis tasks can be performed in the cloud. As such, the computing load including feature extraction and analysis is imposed on the cloud side. \label{fig:trans_a}]
{\includegraphics[width=0.5\textwidth]{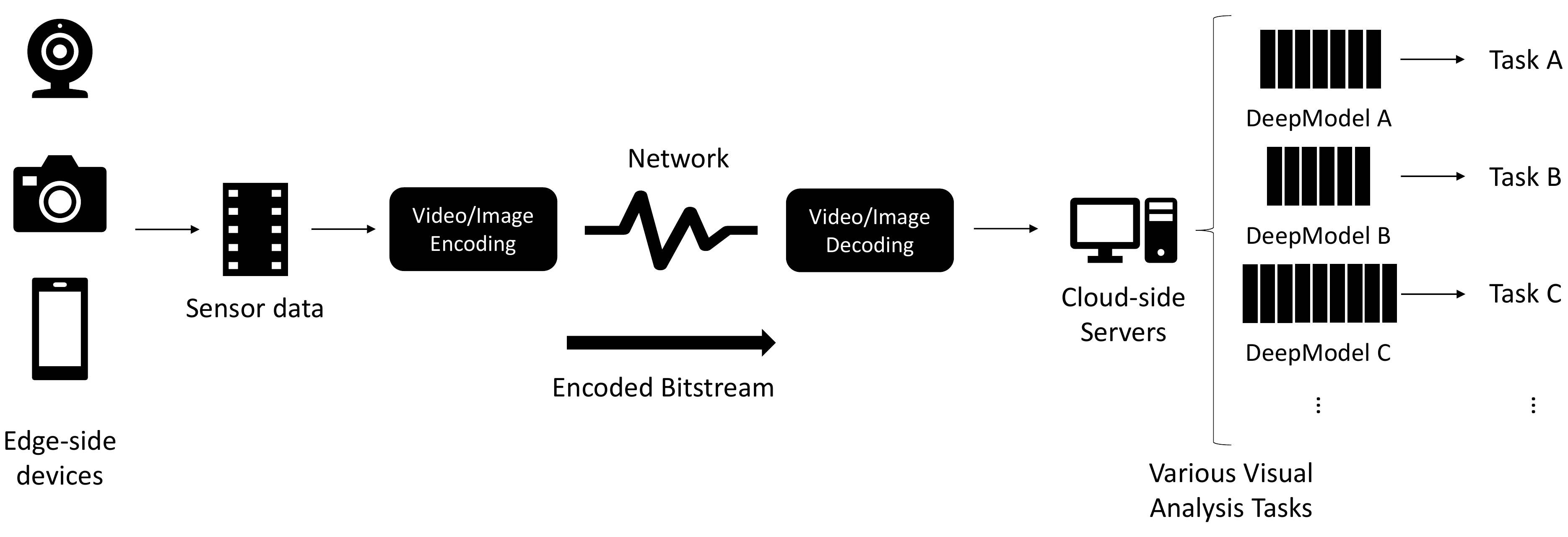}}
\subfigure[Top-layer deep feature transmission. Computing load can be distributed to each front-end device. Only the specific type of analysis is performed corresponding to the deep learning model used at the front-end. \label{fig:trans_b}]
{\includegraphics[width=0.5\textwidth]{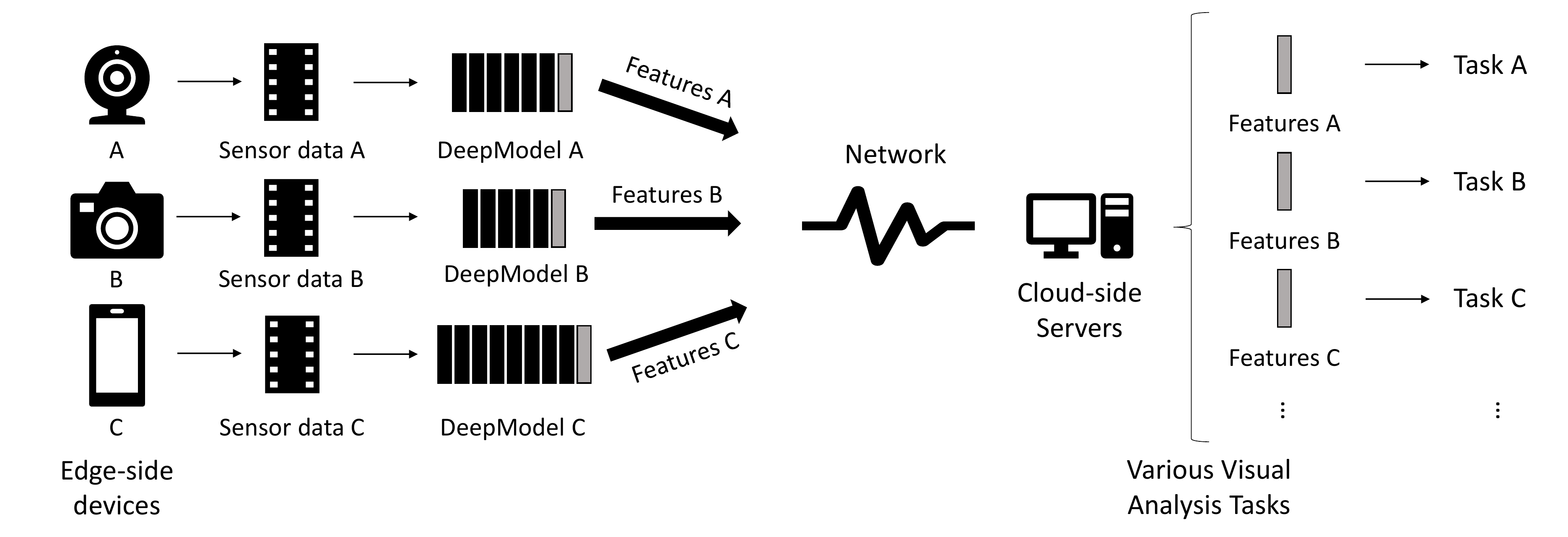}}
\caption{Two commonly used strategies for cloud-based visual analysis.}
\label{fig:trans}
\end{figure}

In cloud-based visual analysis scenarios, visual signal acquisition and analysis are processed in distributed devices. In particular, images/videos are usually acquired in front-end devices (e.g. mobile phones, surveillance cameras) while the analysis is completed in the cloud side. As such, the data transmission between the edge and cloud sides is inevitable. Typically, the data to be transmitted can be either visual signals or features, as shown in Fig.~\ref{fig:trans}.

As the most conventional paradigm, the visual signal compression and transmission methods have been well developed and standardized.  As shown in Fig.~\ref{fig:trans_a}, visual signals (i.e., images and videos) are captured and encoded in the front-end for transmission, and decoded and analyzed in the cloud-end after receiving the bit-stream. More specifically, various analysis tasks can be performed in the cloud-end, since the original visual signals are available. However, it is questionable that whether such visual signal level transmission can efficiently handle the visual big data. Moreover, although the state-of-the-art coding standards such as H.265/HEVC have dramatically improved the coding efficiency, all the computing load for analysis tasks remain on the cloud side. It is almost impossible for the cloud-side servers to timely analyze all the visual signals sent from the edge side devices in the context of visual big data, as the deep learning models are characterized by high computational complexity. 
For instance, the processing speed of a CNN-based object detection model can reach around 50 FPS with a single Titian X GPU in the best case \cite{redmon2018yolov3}, which is the best performance ever reported to our best knowledge. It implies that one state-of-art GPU card can only process two video signal inputs for one single task in real time. As the edge-side cameras can easily proliferate to a larger population, e.g., a smart city can have over one million surveillance cameras installed, a comparable amount of GPUs should be allocated in the cloud side to timely perform the visual analysis, which is unbearable in terms of economic cost and power consuming.

Benefiting from the development of the low-power AI processors \cite{esmaeilzadeh2012neural,guo2017survey,ionica2015movidius}, deep learning models are able to be implemented on front-end devices.
To reduce the computing load on cloud side, an alternative approach is to transmit the features instead of the visual signals, as shown in Fig.~\ref{fig:trans_b}. In this case, features are extracted right after the visual signals being captured in the front-end devices. Then, after the feature transmission, the cloud-end servers can apply visual analysis based on the features received without the visual signal. As the feature extraction usually takes the majority of computing load in a visual analysis application, the cloud-side server only need to handle light computing loads, such as feature comparison, making visual big data analysis feasible.
For handcrafted features, there are quite a few standards defining the feature extraction, compression and transmission, such as the previously mentioned MPEG-CDVS and CDVA \cite{duan2016overview,duan2017compact}. As the feature extraction substantially performs dimensionality reduction on the original visual signals, the features are usually featured with less generalization ability than the visual signals, such that the transmitted features can only be applied to very specific types of tasks. For examples, the features defined by CDVS are more suitable for image retrieval and matching tasks. The deep learning models, which are learned in a data-driven manner, are much more task-specific and the generalization capability is highly concerned in this scenario. 
Considering the deep learning model as one feature extractor, the top layer feature of a deep model is usually extracted as the visual embedding. Comparing with handcrafted features, although the deep leaning features are more expressive and powerful, they still cannot generalize to all the visual analysis tasks. In summary, transmitting the deep learning features can facilitate the shifting of the computing load from the cloud side to the front side which makes visual big data analysis possible. However, the supported analysis tasks that can be achieved on the cloud side are quite limited. In other words, the availability of visual analysis applications on the cloud side is constrained by the models employed in the front-end devices.

\begin{figure*}[t]
\centering
\includegraphics[width=1.0\textwidth]{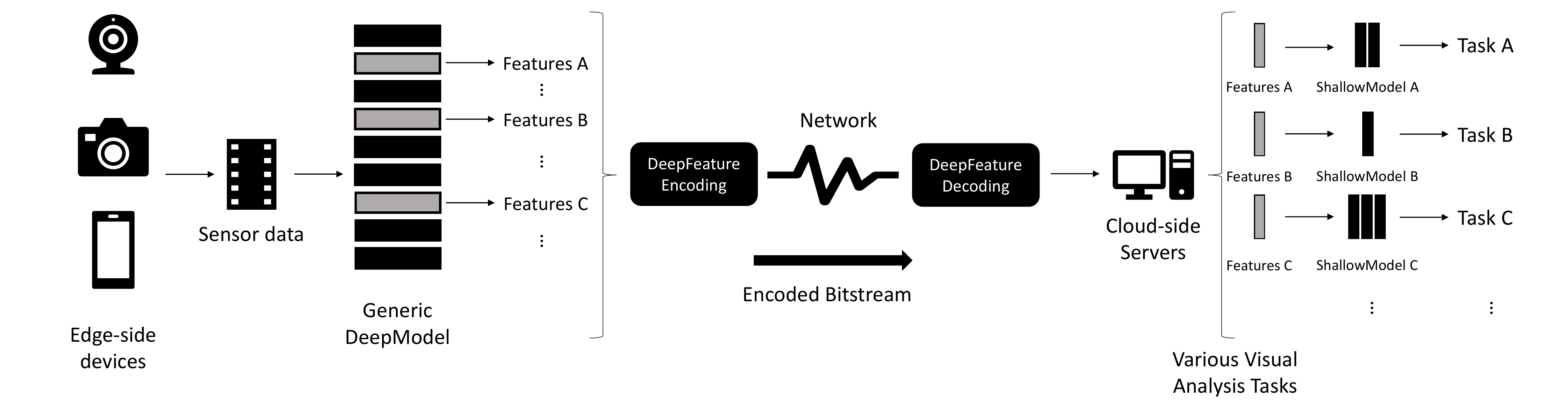}
\vspace{-3 mm}
\caption{Digram of the proposed approach. The intermediate-layer features of a generic deep model can be applied to a broad range of tasks. The features of specific layers will be transmitted based on the analysis requirements on the cloud side. On top of these transmitted features, shallow task-specific models will be applied for visual analysis.}
\label{fig:trans_propose}
\end{figure*}

  Therefore, an approach which can ideally balance the computing load between the front and cloud sides without the limitation of the analysis capability in the cloud side is highly demanded. As shown in Fig.~\ref{fig:trans_propose}, we propose to transmit the intermediate layer features instead of original visual signals and top layer features. Deep learning models are usually characterized by hierarchical structures, which implies that a deep model shall be considered as a combination of stacked feature extractor rather than a single straightforward feature extractor. As such, higher layer features are with large global receptive field which makes them more abstract and task-specific, while lower layer features have a smaller receptive field and location information encoded in 2D feature maps, enabling them to generalize to a broader range of analysis tasks. This provides the flexibility for the cloud side to request appropriate features from the front-end depending on the requirement of analysis task. 

In this case, a generic deep model the features of which can be applied to a broad range of tasks in visual analysis is anticipated to be applied in front-end devices. At present, the commonly used pre-trained CNN models, such as VGGNet and ResNet, which are trained on ImageNet dataset consisting of 1.2 million images of 1000 classes, can be regarded as generic. Features of these deep learning models are wildly adopted in many applications as visual feature extractors. 
For instances, in image captioning tasks, the work \cite{xu2015show} leverages the $conv4$ features (feature map of the fourth convolutional layer, we use the shorthands for convenience in the rest of this paper) of VGGNet to represent given images. The authors in \cite{gu2017stack} encoded the full image with the ResNet to extract both spatial and semantic information from those $conv4$ layer. 
In visual tracking tasks, $conv4$ and $conv5$ features of VGGNet are employed in \cite{wang2015visual}.
In image object detection tasks, the work in \cite{girshick2016region} used the $fc2$ features and $pool5$ features of VGGNet is employed in \cite{girshick2015fast,ren2015faster}.
In visual retrieval, the $pool5$ features of VGGNet are modified to introduce translation, scale and rotation invariances for image retrieval~\cite{lin2017hnip}. Handcrafted features and $fc1$ features of VGGNet are combined to achieve better retrieval performance~\cite{chandrasekhar2016practical}.
In image QA tasks, the $conv5$ features of ResNet is leveraged as the visual representation~\cite{fukui2016multimodal}, and  $pool5$ features of VGGNet and $conv5$ features of ResNet are used in \cite{lu2016hierarchical}.
In view of this, a plenty of visual analysis problems can be solved by applying task-specific neural networks on top of the features extracted by a generic deep model. As the generic model can provide the task-specific neural network with strong representations of the visual signals, a shallow architecture is usually adequate to handle the rest of the visual analysis task which is favorable in terms of the computing costs. Thus, the deployment of our proposed data transmission approach can minimize the computing load on the cloud side while maximizing the availability of various analysis types. Furthermore, it is envisioned that in the future the deep learning models will be developed to more and more generic. At that stage, our proposed approach will have more advantages over the former ones. 


\section{Deep learning feature compression}
\label{sec:feature_compression}

Transmitting intermediate-layer features instead of top-layer features and visual signals is superior at easing the computing load of the cloud end and maintaining the availability of various analysis tasks. However, the transmission bandwidth may limit the deployment of such approach, as the data volume of the intermediate-layer features is non-negligible. In deep learning models, the feature volume of first few layers can be even larger than the input visual signals, as shown in Table~\ref{tab:arch}. 
As such, to optimize the bandwidth, compression for deep features is necessary.


\subsection{Features of Deep Neural Networks}
\label{sec:deepfeature}
As the deep neural networks are characterized with a hierarchical structure of multiple layers, a group of features can be extracted, where the outputs of each layer of the deep model can be considered as features. In the rest of this section, features of convolutional neural networks (CNNs), which are the dominant deep model type in the visual computing task, will be investigated.

Typically, CNN consists of convolutional layers, normalization layers, pooling layers and fully connected layers as its hidden layers.
The convolutional layer is the core building block of a CNN that is account for most of the computational heavy lifting. It applies convolutional filtering to the input, and generates a 3-D matrix with appointed depth. For convenience, the feature of $\textbf{i}-$th convolutional layer is recorded as $conv\textbf{i}$ in the rest of this paper. 
In general, pooling layers are periodically inserted in-between successive convolutional layers to progressively reduce the spatial size of the representations. The pooling layer operates independently on each slice of the convolutional feature and resizes it spatially by combining the outputs of neuron clusters at previous convolutional layer into a single neuron. The output of the $\textbf{i}-$th pooling layer is denoted as $pool\textbf{i}$ feature. It is also worth noting that some architectures use convolutional layers, instead of pooling layers, to down-sample the input matrix by modifying the stride factors of convolutional layers. Fully connected layers are stacked in the top of a CNN to extract high-level semantic information. Such layer applies connections to all neurons in the previous layer with a matrix multiplication followed by a bias offset. The output of a fully connected layer is a 1-D matrix (i.e. a vector) with fixed size. We call the feature of $\textbf{i}-$th fully connected layer as $fc\textbf{i}$.
Various normalization layers, such as local response normalization (LRN), batch normalization (BN), can also be adopted in a CNN. They regularize the network for better performance. Normalization layers are always parameter-free, and they will not change the shapes of input matrices. Such layers cannot bring the features with new semantic meanings. As such, the outputs of normalization layers will not be discussed in the rest of this paper.

Although various CNN architectures have been proposed in recent years, we find that they share common characteristics in terms of hierarchical structures and feature sizes. Table~\ref{tab:arch} lists four milestone CNN architectures in image classification tasks, including AlexNet \cite{krizhevsky2012imagenet}, VggNet \cite{simonyan2014very}, ResNet \cite{he2016deep} and DenseNet \cite{huang2017densely}. With the same input size, these state-of-the-art CNNs extract the features in a hierarchical manner. In the convolutional part, the sizes of feature maps gradually get reduced along with the inference process. It is found to be regular that the feature map size will be halved after one certain block. Such block can be composed of either one single convolutional layer such as in AlexNet, few stacked convolutional layers such as in VggNet, or some more advanced structures like residual or dense connections of several convolutional layers. Along with the size reduction, feature maps usually can represent more high level semantic information in higher layers. When the feature map size becomes small enough, fully connected layers will be followed to convert the visual characteristics to the task-related semantic space, which will largely erase the spatial information in the feature. It can be easily observed that the CNNs share similar numbers of nodes in fully connected layers, such that the fully connected layer features are with similar volume.
Such observations illustrate that CNNs are with analogical hierarchical structures which can provide semblable features.
It is also worth mentioning that most of these benchmark CNNs use ReLU as the activation function, which constrains the numerical distribution of deep features in a similar range. This property is useful for the deep feature compression.


\begin{table*}[t]
\centering
\caption{Architectures of four benchmark deep convolutional neural networks. `op. unit' stands for operation unit, and it can be either a single layer or a combination of multiple layers. `feat. symbol' is the symbol of feature which indicates the specific type of feature. `feat. size' contains the shape and bit size of the feature. It is worth noting that VGGNet, ResNet and DenseNet all have several variants (with different number of layers), but their variants still share the uniformed feature shapes.}
\label{tab:arch}
\resizebox{\textwidth}{!}{%
\begin{tabular}{c|ccc|ccc|ccc|ccc}
  \hline
  \multirow{2}{*}{Blocks} & \multicolumn{3}{c|}{AlexNet} & \multicolumn{3}{c|}{VGGNet} & \multicolumn{3}{c|}{ResNet} & \multicolumn{3}{c}{DenseNet} \\ \cline{2-13}
   & op. unit & feat. symbol & feat. size & op. unit & feat. symbol & feat. size & op. unit & feat. symbol & feat. size & op. unit & feat. symbol & feat. size \\ \hline
  Input & \multicolumn{12}{c}{$224 \times 224 \times 3$ RGB image} \\ \hline
  \multirow{2}{*}{Conv. Block 1} & single conv & $conv1$ & $56 \times 56 \times 96$ & stacked conv & $conv1$ & $224 \times 224 \times 64$ & \multirow{2}{*}{single conv} & \multirow{2}{*}{$conv1$} & \multirow{2}{*}{$112 \times 112 \times 64$} & \multirow{2}{*}{single conv} & \multirow{2}{*}{$conv1$} & \multirow{2}{*}{$112 \times 112 \times 64$} \\ \cline{2-7}
   & max pool & $pool1$ & $28 \times 28 \times 96$ & max pool & $pool1$ & $112 \times 112 \times 64$ & & & & & & \\ \hline
  \multirow{2}{*}{Conv. Block 2} & single conv & $conv2$ & $28 \times 28 \times 256$ & stacked conv & $conv2$ & $112 \times 112 \times 128$ & max pool & $pool2$ & $56 \times 56 \times 64$ & max pool & $pool2$ & $56 \times 56 \times 64$ \\ \cline{2-13}
   & max pool & $pool2$ & $14 \times 14 \times 256$ & max pool & $pool2$ & $56 \times 56 \times 128$ & residual blk. & $conv2$ & $56 \times 56 \times 256$ & dense + trans. & $conv2$ & $56 \times 56 \times 64$ \\ \hline
  \multirow{2}{*}{Conv. Block 3} & \multirow{2}{*}{single conv} & \multirow{2}{*}{$conv3$} & \multirow{2}{*}{$14 \times 14 \times 384$} & stacked conv & $conv3$ & $56 \times 56 \times 256$ & \multirow{2}{*}{residual blk.} & \multirow{2}{*}{$conv3$} & \multirow{2}{*}{$28 \times 28 \times 512$} & ave. pool & $pool3$ & $28 \times 28 \times 64$ \\ \cline{5-7} \cline{11-13}
   & & & & max pool & $pool3$ & $28 \times 28 \times 256$ & & & & dense + trans. & $conv3$ & $28 \times 28 \times 64$ \\ \hline
  \multirow{2}{*}{Conv. Block 4} & \multirow{2}{*}{single conv} & \multirow{2}{*}{$conv4$} & \multirow{2}{*}{$14 \times 14 \times 384$} & stacked conv & $conv4$ & $28 \times 28 \times 512$ & \multirow{2}{*}{residual blk.} & \multirow{2}{*}{$conv4$} & \multirow{2}{*}{$14 \times 14 \times 1024$} & ave. pool & $pool4$ & $14 \times 14 \times 64$ \\ \cline{5-7} \cline{11-13}
   & & & & max pool & $pool1$ & $14 \times 14 \times 512$ & & & & dense + trans. & $conv4$ & $14 \times 14 \times 64$ \\ \hline
  \multirow{2}{*}{Conv. Block 5} & single conv & $conv5$ & $14 \times 14 \times 256$ & stacked conv & $conv5$ & $14 \times 14 \times 512$ & \multirow{2}{*}{residual blk.} & \multirow{2}{*}{$conv5$} & \multirow{2}{*}{$7 \times 7 \times 2048$} & ave. pool & $pool5$ & $7 \times 7 \times 64$ \\ \cline{5-7} \cline{11-13}
   & max pool & $pool5$ & $7 \times 7 \times 256$ & max pool & $pool5$ & $7 \times 7 \times 512$ & & & & dense & $conv5$ & $7 \times 7 \times ?$ \\ \hline
  \multirow{3}{*}{FC Block} & single fc & $fc1$ & $4096$ & single fc & $fc1$ & $4096$ & \multirow{2}{*}{ave. pool} & \multirow{2}{*}{$pool6$} & \multirow{2}{*}{$1 \times 1 \times 2048$} & \multirow{2}{*}{ave. pool} & \multirow{2}{*}{$pool6$} & \multirow{2}{*}{$1 \times 1 \times ?$} \\ \cline{2-7}
   & single fc & $fc2$ & $4096$ & single fc & $fc2$ & $4096$ & & & & & & \\ \cline{2-13}
   & single fc & $fc3$ & $1000$ & single fc & $fc3$ & $1000$ & single fc & $fc1$ & $1000$ & single fc & $fc1$ & $1000$ \\
  \hline
\end{tabular}
}
\end{table*}

\subsection{Toward Standardization of Deep Feature Compression}
\label{sec:standard}

To ensure compatibility and facilitate interoperability, a series of standards have been established for transmitting visual signal and handcrafted features, as mentioned in Section~\ref{sec:related}. It is envisioned that the proposed approach of compressing the intermediate layer feature is also expected to be standardized in the future.

Different from traditional image/video coding, the conventional feature coding pipeline involves both feature extraction and compression. This is because, for video coding, the source visual signals are well established and available, i.e. the pixel values. By contrast, different visual models would provide different features for the subsequent compression process in feature coding \cite{duan2017ai}. In view of this, to fully ensure the interoperability, both feature extraction and compression process should be specified in the standard for feature coding. In such manner, handcrafted feature coding standards including CDVS and CDVA, carefully standardize the compression and feature extraction methods. Such standardization requires the deterministic method and model. In particular, for deep feature coding, the architecture and parameters of the deep learning model need to be settled.
However, in the current stage, it is different to specify a universal generic deep model which can be applied to all different visual analysis tasks. Furthermore, new deep learning techniques and architectures emerges endlessly. Therefore, the standardization of deep learning model is not ready for prime time. 

Though the deep learning models are kaleidoscopic, the deep features share similar shapes and distributions in specific layers as discussed in Section~\ref{sec:deepfeature}. Based on this observation, only the feature compression needs to be standardized. In other words, only the pipeline from raw deep features to the compressed bitstream is taken into consideration, and the final syntax that specifies the deep features is standardized. In this way, the choice of deep model is left open for the system customization. In the future, any effective deep learning models can seamlessly collaborate with this standard, which help the standard keep with long-lasting vitality. With this standardization strategy, the explosion of deep learning techniques and the interoperability can be well balanced.

Regarding the deep feature compression, it is expected to remove the redundancy of deep learning features in both single images and video sequences. Also, deep feature compression methods should be either lossless or lossy. This is very similar to video coding standards such as HEVC which supports both image/video compression and lossless/lossy methods. Instead of being uniform as the image or video signals, the characteristics of deep features are more diverging. For example, features of convolutional layers are in the form of feature maps which is very different from features of fully-connected layers that are in vectors. On account of this, there should be different compression strategies for distinct feature categories (i.e. $conv$, $pool$, $fc$). For the $conv$ and $pool$ feature which is a combination of spatial 2D signals, many video coding technologies can be transferred to the feature coding, such as inter prediction, intra prediction and rate distortion optimization. For the $fc$ feature which is a vector, general data compression methods can be referred, such as entropy coding. As the dynamic range of deep feature values is commonly smaller comparing with the numerical range of its data type, quantization methods should be efficient to remove the redundancy. As such, how the redundancies of deep features can be removed and how to minimize the performance drop of deep feature while maximizing the redundancy reduction should be further investigated in the standardization exploration.

\section{Evaluations on Lossless Compression of Intermediate Deep Learning Features }
\label{sec:evaluation}
In this section, we present the evaluation results of the lossless compression of intermediate deep learning features.
By evaluating the benchmark lossless data compression methods on deep learning features extracted with several wildly-used networks, we aim to provide the baselines for further research and standardization activities.

\subsection{Experiment Setup}

To provide the meaningful baseline evaluations, we carefully selected the generic deep learning models and data compression methods.
In particular, the deep learning models are chosen based on the principle that the extracted intermediate features should be generic enough to be applied to a wide range of tasks in visual analysis. Then several conventional and wildly used compression algorithms are selected to perform deep feature compression.

\subsubsection{Deep Learning Models and Datasets}
\label{sec:generic_models}

In this paper, we adopt official models of VGGNets and ResNets to perform feature extraction. 
These commonly used pre-trained CNN models are the winners of ImageNet Large Scale Visual Recognition Challenge (ILSVRC) 2014 and 2015, which are trained on ImageNet dataset consisting of 1.2 million images of 1000 classes, such that the features can be regarded as generic.
As such, the bellwethers of ILSVRC become the common choice for image feature extraction. 
As mentioned in Section~\ref{sec:transmit_feature}, for many computer vision applications, the task-specific models are designed on top of the features of VGGNets and ResNets, such as image captioning \cite{xu2015show,gu2017stack}, visual tracking~\cite{wang2015visual}, image object detection~\cite{girshick2016region,girshick2015fast,ren2015faster,redmon2018yolov3}, visual retrieval~\cite{lin2017hnip,chandrasekhar2016practical}, image QA~\cite{fukui2016multimodal,lu2016hierarchical}.

\par{VGGNet:}
Simonyan and Zisserman developped VGGNet at the ILSVRC 2014. VGG-16 outstands from the six variants of VGGNet for its good balance among performance and computational complexity. VGG-16 consists of 16 convolutional layers and is very appealing because of its neat architecture. It only performs $3\times 3$ convolutions and $2\times 2$ pooling all the way through. Currently it is the most preferred choice in the community for extracting features from images. 

\par{ResNet:}
At the ILSVRC 2015, He \textit{et al.} introduced Residual Neural Network (ResNet) which contains a novel technique called ``skip connections''. Thanks to this new structure, the networks are able to go into very deep (152 layers in He \textit{et al.}'s work) with lower complexity than VGGNet. ResNets have three commonly used variants with 50, 101, 152 layers respectively. Benefited from the astonishing performance (top-5 error rate of 5.25\%, 4.60\%, 4.49\%), ResNets are increasingly adopted by various tasks. 

We extract the deep features of the aforementioned deep learning models on a subset of the validation set of the ImageNet 2012 dataset \cite{russakovsky2015imagenet}. To economic the subsequent compression time while maintaining the variety of test image categories, we randomly choose one image from each of the 1,000 classes.
Overall, we evaluate the compression performance on each feature type with 1,000 feature entities.

\subsubsection{Compression Methods}
\label{sec:compress_methods}

In analogous to data compression, deep feature compression aims to encode deep learning features with fewer bits than the original, which can be either lossless or lossy. The lossless compression ensures that the decoded feature is identical with the one before encoding. As such, analysis performance degradation will not be introduced. Lossy feature compression reduces the data size by eliminating less important information, which may end in performance loss of corresponding deep learning models.
In this paper, we evaluate the performance with four conventional lossless data compression methods, and lossy methods will be explored in the future studies. The adopted compression methods are described as follows,

\par{GZIP:}
GZIP \cite{GZIP} was developed in the early 1990's as a replacement for patent-encumbered algorithms such as LZW \cite{welch1984technique}. The DEFLATE algorithm \cite{deutsch1996deflate} is the core of GZIP, which employs LZ77 \cite{ziv1977universal} followed by Huffman coding \cite{huffman1952method}. The GZIP algorithm enjoys very fast compression speed and small memory footprint.

\par{ZLIB:}
ZLIB \cite{ZLIB} was adapted from the GZIP in mid-1990's. It abstracts the DEFLATE algorithm to achieve higher compression ratio and faster speed. ZLIB is now widely used for data transmission and storage.

\par{BZIP2:}
BZIP2 \cite{BZIP2} compresses the initial data with Run-length encoding (RLE) and applies the Burrows-Wheeler transform to rearrange character strings into runs of similar characters. It then uses move-to-front (MTF) transform and a combination of RLE and Huffman coding to efficiently represent the data stream.
BZIP2 is generally considered with higher compression ratio than the LZW and Deflate algorithms with relatively slower speed.

\par{LZMA:}
The Lempel–Ziv–Markov chain algorithm (LZMA) \cite{LZMA}  uses a dictionary compression scheme, similar to LZ77 \cite{ziv1977universal}, followed by a range encoder. Comparing to LZ77, the dictionary compressor is with huge dictionary sizes (up to 4 GB). The range encoder employs a complex mechanism to make probability predictions of each bit. LZMA features a generally high compression ratio with a comparable speed~\cite{collin2005quick}.

\subsection{Results}
\label{sec:experiment}
To evaluate the deep feature compression performance, deep features are firstly extracted from different layers of deep models on the aforementioned subset of ImageNet dataset. Subsequently, four classic lossless compression algorithms with default configurations are applied on the extracted features.
The feature extractions are performed by Caffe and Tensorflow on a NVIDIA GeForce 1080 GPU. The compression processes are conducted on Intel Xeon CPU E5-2650 v2 @ 2.60GHz with only one thread.

We mainly consider two criteria to evaluate the compression performance: compression rate and computational time cost. In particular, the compression rate is defined as
\begin{equation}
    Compression\ rate = \frac{data\ length\ after\ compression}{data\ length\ before\ compression}.
\end{equation}
In this paper, we report the mean compression rate and computational time over 1,000 samples of each type of the deep learning features for the four lossless compression methods.
The statistics of each type of features, such as the shape, volume and non-zero rate, are also demonstrated.
In particular, from the observation that ReLU function in deep learning models can result in a plenty of identical values (i.e. zeros), which may directly affect the compression rate, we list the mean non-zero rates of each type of feature for compression rates comparison. The results are listed in Tables~\ref{tab:vgg} to~\ref{tab:res152} for VGGNet-16, ResNet-50, ResNet-101, ResNet-152, respectively.

\begin{table*}
  \centering
  \caption{Feature compression results of VGGNet.}
  \label{tab:vgg}
  \resizebox{\textwidth}{!}{
  \begin{tabular}{cccc|cc|cc|cc|cc}
  \hline
    \multirow{2}{*}{Feat. Type} & \multirow{2}{*}{Feat. Shape} & \multirow{2}{*}{Data Volume} & \multirow{2}{*}{Non-zero} & \multicolumn{2}{c}{GZIP} & \multicolumn{2}{c}{ZLIB} & \multicolumn{2}{c}{BZIP2} & \multicolumn{2}{c}{LZMA}\\
    \cline{5-12}
     & & & & Comp. Ratio & Time Cost & Comp. Ratio & Time Cost & Comp. Ratio & Time Cost & Comp. Ratio & Time Cost\\
    \hline
    $conv1$ & $224\times 224\times 64$ & $12.25M$ & $0.685\pm 0.021$ & $0.639\pm 0.044$ & $2.016\pm 0.345$ & $0.641\pm 0.044$ & $0.521\pm 0.040$ & $0.648\pm 0.044$ & $3.170\pm 0.572$ & $0.609\pm 0.045$ & $5.146\pm 0.585$\\
    $pool1$ & $112\times 112\times 64$ & $3.0625M$ & $0.815\pm 0.023$ & $0.752\pm 0.055$ & $0.345\pm 0.050$ & $0.753\pm 0.055$ & $0.147\pm 0.013$ & $0.766\pm 0.053$ & $0.688\pm 0.227$ & $0.719\pm 0.058$ & $0.815\pm 0.056$\\
    $conv2$ & $112\times 112\times 128$ & $6.125M$ & $0.480\pm 0.012$ & $0.482\pm 0.028$ & $1.369\pm 0.199$ & $0.486\pm 0.028$ & $0.216\pm 0.015$ & $0.475\pm 0.026$ & $1.556\pm 0.489$ & $0.460\pm 0.028$ & $2.451\pm 0.330$\\
    $pool2$ & $56\times 56\times 128$ & $1568K$ & $0.694\pm 0.033$ & $0.680\pm 0.046$ & $0.378\pm 0.039$ & $0.683\pm 0.046$ & $0.075\pm 0.006$ & $0.672\pm 0.043$ & $0.239\pm 0.076$ & $0.655\pm 0.046$ & $0.493\pm 0.021$\\
    $conv3$ & $56\times 56\times 256$ & $3.0625M$ & $0.302\pm 0.026$ & $0.319\pm 0.030$ & $0.541\pm 0.098$ & $0.322\pm 0.030$ & $0.077\pm 0.007$ & $0.308\pm 0.028$ & $1.188\pm 0.095$ & $0.301\pm 0.028$ & $1.130\pm 0.120$\\
    $pool3$ & $28\times 28\times 256$ & $784K$ & $0.484\pm 0.049$ & $0.502\pm 0.050$ & $0.241\pm 0.032$ & $0.506\pm 0.050$ & $0.031\pm 0.003$ & $0.484\pm 0.047$ & $0.139\pm 0.055$ & $0.478\pm 0.049$ & $0.259\pm 0.021$\\
    $conv4$ & $28\times 28\times 512$ & $1568K$ & $0.127\pm 0.014$ & $0.146\pm 0.016$ & $0.124\pm 0.016$ & $0.148\pm 0.016$ & $0.023\pm 0.002$ & $0.138\pm 0.015$ & $0.511\pm 0.022$ & $0.137\pm 0.014$ & $0.348\pm 0.054$\\
    $pool4$ & $14\times 14\times 512$ & $392K$ & $0.243\pm 0.030$ & $0.274\pm 0.033$ & $0.071\pm 0.011$ & $0.278\pm 0.033$ & $9.969m\pm 1.114m$ & $0.259\pm 0.030$ & $0.131\pm 0.019$ & $0.255\pm 0.030$ & $0.122\pm 0.012$\\
    $conv5$ & $14\times 14\times 512$ & $392K$ & $0.068\pm 0.017$ & $0.079\pm 0.018$ & $0.024\pm 0.004$ & $0.081\pm 0.019$ & $4.045m\pm 0.411m$ & $0.075\pm 0.018$ & $0.108\pm 0.002$ & $0.074\pm 0.017$ & $0.048\pm 0.010$\\
    $pool5$ & $7\times 7\times 512$ & $98K$ & $0.124\pm 0.028$ & $0.147\pm 0.031$ & $0.013\pm 0.002$ & $0.150\pm 0.032$ & $1.498m\pm 0.213m$ & $0.143\pm 0.030$ & $0.029\pm 0.001$ & $0.139\pm 0.029$ & $0.018\pm 0.003$\\
    $fc1$ & $4096\times 1$ & $16K$ & $0.248\pm 0.046$ & $0.304\pm 0.047$ & $5.204m\pm 0.662m$ & $0.307\pm 0.047$ & $0.537m\pm 0.062m$ & $0.305\pm 0.046$ & $2.918m\pm 0.191m$ & $0.288\pm 0.045$ & $6.800m\pm 0.849m$\\
    $fc2$ & $4096\times 1$ & $16K$ & $0.259\pm 0.061$ & $0.315\pm 0.062$ & $4.964m\pm 0.485m$ & $0.318\pm 0.062$ & $0.540m\pm 0.070m$ & $0.317\pm 0.061$ & $2.933m\pm 0.191m$ & $0.300\pm 0.060$ & $6.257m\pm 0.249m$\\
    $fc3$ & $1000\times 1$ & $4000$ & $1.000\pm 0.000$ & $0.940\pm 0.002$ & $0.156m\pm 0.016m$ & $0.937\pm 0.002$ & $0.115m\pm 0.008m$ & $1.086\pm 0.004$ & $1.310m\pm 0.009m$ & $0.964\pm 0.006$ & $2.007m\pm 0.090m$\\
  \hline
\end{tabular}
}
\end{table*}

\begin{table*}
  \centering
  \caption{Feature compression results of ResNet-50.}
  \label{tab:res50}
  \resizebox{\textwidth}{!}{
  \begin{tabular}{cccc|cc|cc|cc|cc}
  \hline
    \multirow{2}{*}{Feat. Type} & \multirow{2}{*}{Feat. Shape} & \multirow{2}{*}{Data Volume} & \multirow{2}{*}{Non-zero} & \multicolumn{2}{c}{GZIP} & \multicolumn{2}{c}{ZLIB} & \multicolumn{2}{c}{BZIP2} & \multicolumn{2}{c}{LZMA}\\
    \cline{5-12}
     & & & & Comp. Ratio & Time Cost & Comp. Ratio & Time Cost & Comp. Ratio & Time Cost & Comp. Ratio & Time Cost\\
    \hline
    $conv1$ & $112\times 112\times 64$ & $3.0625M$ & $0.686\pm 0.026$ & $0.619\pm 0.019$ & $0.254\pm 0.091$ & $0.619\pm 0.019$ & $0.117\pm 0.006$ & $0.639\pm 0.021$ & $1.067\pm 0.047$ & $0.578\pm 0.023$ & $0.765\pm 0.083$\\
    $pool1$ & $56\times 56\times 64$ & $784K$ & $0.765\pm 0.031$ & $0.529\pm 0.025$ & $0.050\pm 0.009$ & $0.529\pm 0.025$ & $0.024\pm 0.001$ & $0.638\pm 0.029$ & $0.261\pm 0.017$ & $0.459\pm 0.024$ & $0.162\pm 0.014$\\
    $conv2$ & $56\times 56\times 256$ & $3.0625M$ & $0.707\pm 0.028$ & $0.681\pm 0.018$ & $0.595\pm 0.089$ & $0.683\pm 0.018$ & $0.138\pm 0.003$ & $0.681\pm 0.022$ & $0.442\pm 0.069$ & $0.652\pm 0.016$ & $1.067\pm 0.091$\\
    $conv3$ & $28\times 28\times 512$ & $1568K$ & $0.686\pm 0.014$ & $0.672\pm 0.011$ & $0.327\pm 0.037$ & $0.674\pm 0.011$ & $0.068\pm 0.002$ & $0.666\pm 0.012$ & $0.204\pm 0.021$ & $0.649\pm 0.011$ & $0.488\pm 0.049$\\
    $conv4$ & $14\times 14\times 1024$ & $784K$ & $0.541\pm 0.024$ & $0.548\pm 0.021$ & $0.172\pm 0.017$ & $0.550\pm 0.021$ & $0.030\pm 0.001$ & $0.535\pm 0.022$ & $0.126\pm 0.031$ & $0.526\pm 0.021$ & $0.238\pm 0.013$\\
    $conv5$ & $7\times 7\times 2048$ & $392K$ & $0.176\pm 0.032$ & $0.200\pm 0.033$ & $0.065\pm 0.011$ & $0.203\pm 0.034$ & $7.311m\pm 0.943m$ & $0.190\pm 0.032$ & $0.115\pm 0.006$ & $0.188\pm 0.032$ & $0.086\pm 0.012$\\
    $pool5$ & $1\times 1\times 2048$ & $8K$ & $0.887\pm 0.063$ & $0.858\pm 0.041$ & $0.405m\pm 0.176m$ & $0.857\pm 0.041$ & $0.265m\pm 0.035m$ & $0.935\pm 0.053$ & $1.852m\pm 0.048m$ & $0.861\pm 0.044$ & $2.476m\pm 0.081m$\\
    $fc1$ & $1000\times 1$ & $4000$ & $1.000\pm 0.000$ & $0.941\pm 0.002$ & $0.155m\pm 0.015m$ & $0.938\pm 0.002$ & $0.115m\pm 0.012m$ & $1.086\pm 0.004$ & $1.246m\pm 0.021m$ & $0.965\pm 0.006$ & $1.699m\pm 0.038m$\\
  \hline
\end{tabular}
}
\end{table*}

\begin{table*}
  \centering
  \caption{Feature compression results of ResNet-101.}
  \label{tab:res101}
  \resizebox{\textwidth}{!}{
  \begin{tabular}{cccc|cc|cc|cc|cc}
  \hline
    \multirow{2}{*}{Feat. Type} & \multirow{2}{*}{Feat. Shape} & \multirow{2}{*}{Data Volume} & \multirow{2}{*}{Non-zero} & \multicolumn{2}{c}{GZIP} & \multicolumn{2}{c}{ZLIB} & \multicolumn{2}{c}{BZIP2} & \multicolumn{2}{c}{LZMA}\\
    \cline{5-12}
     & & & & Comp. Ratio & Time Cost & Comp. Ratio & Time Cost & Comp. Ratio & Time Cost & Comp. Ratio & Time Cost\\
    \hline
    $conv1$ & $112\times 112\times 64$ & $3.0625M$ & $0.660\pm 0.023$ & $0.588\pm 0.017$ & $0.216\pm 0.070$ & $0.588\pm 0.018$ & $0.110\pm 0.005$ & $0.610\pm 0.020$ & $1.042\pm 0.031$ & $0.543\pm 0.021$ & $0.640\pm 0.047$\\
    $pool1$ & $56\times 56\times 64$ & $784K$ & $0.713\pm 0.026$ & $0.489\pm 0.023$ & $0.041\pm 0.007$ & $0.490\pm 0.023$ & $0.022\pm 0.001$ & $0.590\pm 0.026$ & $0.261\pm 0.008$ & $0.422\pm 0.022$ & $0.143\pm 0.015$\\
    $conv2$ & $56\times 56\times 256$ & $3.0625M$ & $0.687\pm 0.037$ & $0.663\pm 0.026$ & $0.601\pm 0.091$ & $0.665\pm 0.026$ & $0.131\pm 0.004$ & $0.662\pm 0.030$ & $0.441\pm 0.059$ & $0.632\pm 0.024$ & $0.980\pm 0.067$\\
    $conv3$ & $28\times 28\times 512$ & $1568K$ & $0.724\pm 0.021$ & $0.703\pm 0.015$ & $0.325\pm 0.042$ & $0.705\pm 0.014$ & $0.070\pm 0.002$ & $0.699\pm 0.017$ & $0.201\pm 0.011$ & $0.676\pm 0.013$ & $0.427\pm 0.017$\\
    $conv4$ & $14\times 14\times 1024$ & $784K$ & $0.730\pm 0.032$ & $0.711\pm 0.025$ & $0.180\pm 0.037$ & $0.714\pm 0.024$ & $0.036\pm 0.002$ & $0.704\pm 0.028$ & $0.100\pm 0.005$ & $0.691\pm 0.025$ & $0.202\pm 0.009$\\
    $conv5$ & $7\times 7\times 2048$ & $392K$ & $0.164\pm 0.035$ & $0.187\pm 0.037$ & $0.060\pm 0.011$ & $0.190\pm 0.037$ & $6.516m\pm 0.957m$ & $0.178\pm 0.035$ & $0.116\pm 0.006$ & $0.176\pm 0.035$ & $0.079\pm 0.013$\\
    $pool5$ & $1\times 1\times 2048$ & $8K$ & $0.860\pm 0.075$ & $0.841\pm 0.051$ & $0.455m\pm 0.214m$ & $0.840\pm 0.051$ & $0.276m\pm 0.035m$ & $0.914\pm 0.064$ & $1.848m\pm 0.062m$ & $0.843\pm 0.054$ & $2.544m\pm 0.104m$\\
    $fc1$ & $1000\times 1$ & $4000$ & $1.000\pm 0.000$ & $0.941\pm 0.002$ & $0.146m\pm 0.005m$ & $0.938\pm 0.002$ & $0.115m\pm 0.008m$ & $1.086\pm 0.004$ & $1.233m\pm 0.030m$ & $0.965\pm 0.006$ & $1.746m\pm 0.039m$\\
  \hline
\end{tabular}
}
\end{table*}

\begin{table*}
  \centering
  \caption{Feature compression results of ResNet-152.}
  \label{tab:res152}
  \resizebox{\textwidth}{!}{
  \begin{tabular}{cccc|cc|cc|cc|cc}
  \hline
    \multirow{2}{*}{Feat. Type} & \multirow{2}{*}{Feat. Shape} & \multirow{2}{*}{Data Volume} & \multirow{2}{*}{Non-zero} & \multicolumn{2}{c}{GZIP} & \multicolumn{2}{c}{ZLIB} & \multicolumn{2}{c}{BZIP2} & \multicolumn{2}{c}{LZMA}\\
    \cline{5-12}
     & & & & Comp. Ratio & Time Cost & Comp. Ratio & Time Cost & Comp. Ratio & Time Cost & Comp. Ratio & Time Cost\\
    \hline
    $conv1$ & $112\times 112\times 64$ & $3.0625M$ & $0.595\pm 0.026$ & $0.527\pm 0.022$ & $0.177\pm 0.046$ & $0.527\pm 0.022$ & $0.102\pm 0.005$ & $0.548\pm 0.025$ & $0.981\pm 0.049$ & $0.485\pm 0.024$ & $0.679\pm 0.057$\\
    $pool1$ & $56\times 56\times 64$ & $784K$ & $0.640\pm 0.029$ & $0.441\pm 0.025$ & $0.040\pm 0.008$ & $0.442\pm 0.025$ & $0.020\pm 0.001$ & $0.532\pm 0.029$ & $0.248\pm 0.010$ & $0.379\pm 0.023$ & $0.144\pm 0.012$\\
    $conv2$ & $56\times 56\times 256$ & $3.0625M$ & $0.715\pm 0.030$ & $0.681\pm 0.021$ & $0.538\pm 0.076$ & $0.682\pm 0.021$ & $0.138\pm 0.006$ & $0.685\pm 0.024$ & $0.683\pm 0.186$ & $0.644\pm 0.017$ & $1.121\pm 0.096$\\
    $conv3$ & $28\times 28\times 512$ & $1568K$ & $0.757\pm 0.023$ & $0.729\pm 0.017$ & $0.322\pm 0.048$ & $0.731\pm 0.017$ & $0.072\pm 0.003$ & $0.727\pm 0.020$ & $0.201\pm 0.011$ & $0.704\pm 0.016$ & $0.485\pm 0.032$\\
    $conv4$ & $14\times 14\times 1024$ & $784K$ & $0.765\pm 0.031$ & $0.740\pm 0.023$ & $0.165\pm 0.038$ & $0.742\pm 0.022$ & $0.037\pm 0.002$ & $0.735\pm 0.026$ & $0.098\pm 0.004$ & $0.719\pm 0.023$ & $0.231\pm 0.016$\\
    $conv5$ & $7\times 7\times 2048$ & $392K$ & $0.165\pm 0.034$ & $0.188\pm 0.036$ & $0.062\pm 0.011$ & $0.191\pm 0.036$ & $6.835m\pm 1.000m$ & $0.178\pm 0.034$ & $0.121\pm 0.006$ & $0.177\pm 0.034$ & $0.085\pm 0.013$\\
    $pool5$ & $1\times 1\times 2048$ & $8K$ & $0.860\pm 0.074$ & $0.841\pm 0.050$ & $0.469m\pm 0.218m$ & $0.840\pm 0.050$ & $0.276m\pm 0.035m$ & $0.914\pm 0.063$ & $1.980m\pm 0.054m$ & $0.843\pm 0.054$ & $2.841m\pm 0.096$\\
    $fc1$ & $1000\times 1$ & $4000$ & $1.000\pm 0.000$ & $0.941\pm 0.002$ & $0.155m\pm 0.003m$ & $0.938\pm 0.002$ & $0.115m\pm 0.007m$ & $1.086\pm 0.004$ & $1.286m\pm 0.023m$ & $0.965\pm 0.006$ & $1.986m\pm 0.037m$\\
  \hline
\end{tabular}
}
\end{table*}

From the tables we can see that, in terms of compression time cost, the ZLIB method is with the standout compression speed. It takes less time than the other three methods conspicuously on each type of feature. On the contrary, the speed performance of GZIP, BZIP2 and LZMA varies on different feature types. For instance, when compressing the large-volume features (e.g., $conv1$ of VGGNet with 12.25MByte) and small-volume features (e.g., $pool5$ of ResNets and $fc$ features which are under 16KByte), LZMA is the slowest among the four methods. When dealing with features of the volume between 98KByte and 3.0625MByte, BZIP2 takes longer time than LZMA in some cases (e.g., $conv4$-$pool5$ of VGGNet and $conv1$, $pool1$, $conv4$, $conv5$ of ResNet). GZIP takes less time than LZMA and BZIP2 in most cases. However, regarding $pool2$, $pool3$ of VGGNet and $conv2$, $conv3$ of ResNet, which are with the volume between 784KByte and 3.0625MByte, BZIP2 is faster.
Overall, the compression speed of ZLIB is orders of magnitude faster than the other three. GZIP, BZIP2 and LZMA are with comparable time cost, while GZIP is generally faster and LZMA is relatively slow among the three. The speed of BZIP2 is not stable and highly depends on feature types.

\begin{figure}[t]
\centering
\subfigure[Compression rate on VGGNet-16 features. \label{fig:result_vgg}]
{\includegraphics[width=0.5\textwidth]{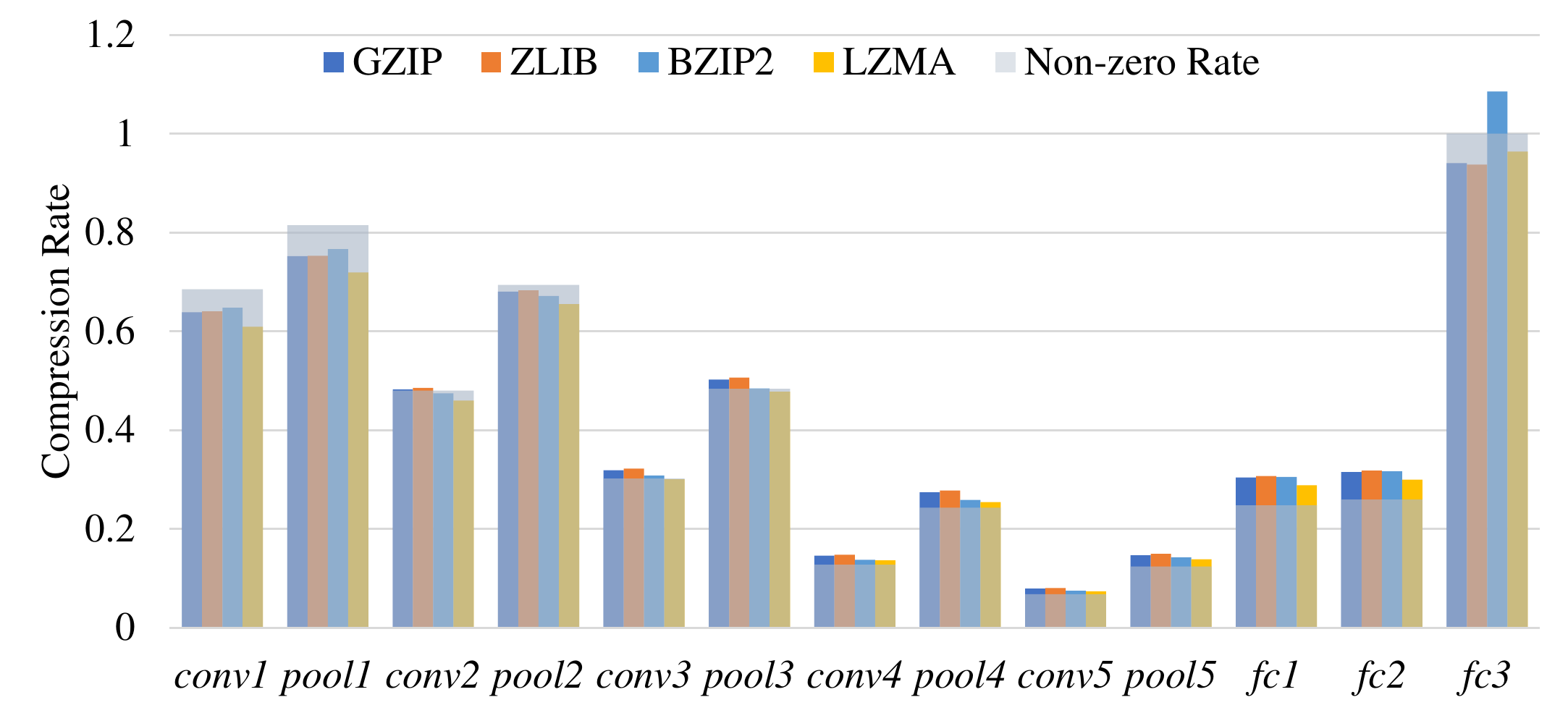}}
\subfigure[Compression rate on ResNet-50 features. \label{fig:result_res50}]
{\includegraphics[width=0.5\textwidth]{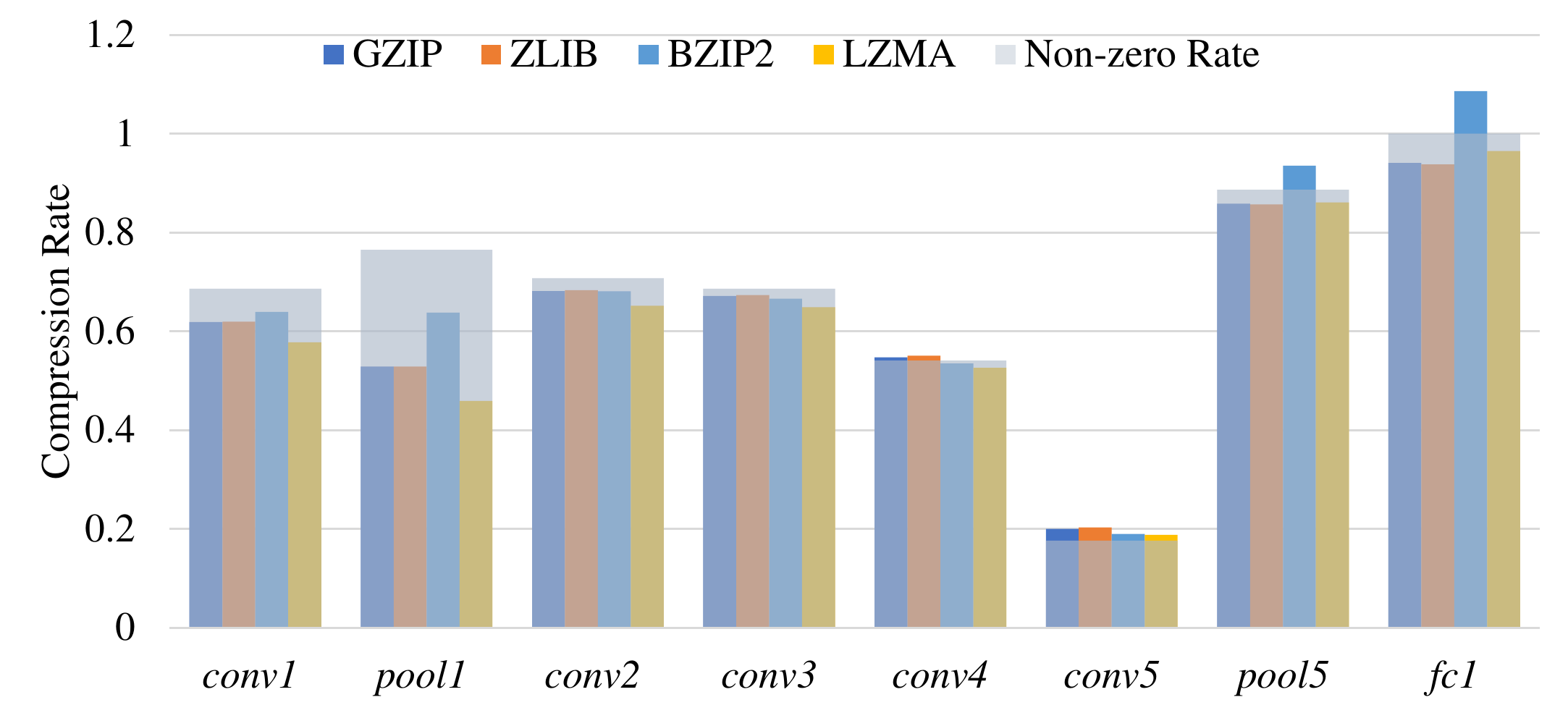}}
\subfigure[Compression rate on ResNet-101 features. \label{fig:result_res101}]
{\includegraphics[width=0.5\textwidth]{figs/fig_results_res50.pdf}}
\subfigure[Compression rate on ResNet-152 features. \label{fig:result_res152}]
{\includegraphics[width=0.5\textwidth]{figs/fig_results_res50.pdf}}
\caption{Compression rate comparison of four benchmark methods. The non-zeros rate of each feature type is shown as the broad gray bar for reference.}
\label{fig:results}
\end{figure}

Regarding the compression performance, the compression rates on different features with different benchmark methods are illustrated in Fig.~\ref{fig:results}.
We can see that the performance of LZMA is superior to the other three methods on most of feature types except for $fc3$ of VGGNet and $pool5$/$fc1$ of ResNets, while ZLIB performs better on $fc3$ of VGGNet and $pool5$/$fc1$ of ResNets. 
GZIP has similar performance compared with ZLIB, though it wins ZLIB a little bit on compressing features except for $fc3$ of VGGNet and $pool5$/$fc1$ of ResNets. 
BZIP2 provides comparable compression rates on features except for $fc3$ of VGGNet and $pool5$/$fc1$ of ResNets, whereas its performance on $fc3$ of VGGNet and $pool5$/$fc1$ of ResNets is dramatically worse than the other three methods. In particular, the compression rates of BZIP2 on the final layer features of all the four tested networks are higher than $1.0$, which implies that the compressed data volume is even larger than the uncompressed one.
The above observations show that the performance of the four methods on $fc3$ of VGGNet and $pool5$/$fc1$ of ResNets is largely different from the other feature types.
This may be because that the distributions of $fc3$ of VGGNet and $pool5$, $fc1$ of ResNets are different from the others. These three types of features are from the top layers of the corresponding neural networks. Unlike the low level layer features which are stacked 2-D maps with remaining spatial correlations among the elements, the top layer features are in the form of 1-D vector which is lack of correlations between its elements as mentioned in Section~\ref{sec:deepfeature}. Furthermore, these three types of features are of higher non-zero rates than the other feature types, which may also affects the performance of the compression methods.
From Fig.~\ref{fig:results}, we can also find that the compression rates of the four compression methods on one feature type are highly related with the non-zero rates of this feature type. It may be because that ReLU functions in neural networks provides a number of zero values in a feature sample, which produce statistical redundancy in the feature. The non-zero elements in the feature usually distribute on a broad numerical range, making the possibility very low to have several elements with the same value. As such, it is difficult for the compression methods to exploit the statistical redundancy in non-zero elements. Therefore, the performance of the lossless compression methods are largely affected by the non-zero rate of a feature sample.

In summary, regarding lossless compression of the deep learning features, the compression rates of the four benchmark data compression methods are around the non-zero rate of the deep feature. LZMA achieves the best compression rates on most of the feature types with the highest computational complexity. ZLIB performs comparably in term of compression rate with much shorter time. GZIP performs well but not the best in terms of both compression rate and computational cost. The performance of BZIP2 is not stable in terms of both compression rate and time cost, and highly depends on the feature type.

\subsection{Discussions}
By evaluating the four benchmark lossless data compression methods on deep learning features, we notice that the compression rate of a lossless compression method is largely limited by the non-zero rate of the feature to be compressed. The statistical redundancy of the deep learning feature mainly depends on the elements with zero value. It is difficult to identify and eliminate statistical redundancy from the non-zero elements of the deep learning features. As such, compressing the deep features in a lossless manner does not guarantee much room to improve. From the evaluation results, the best compression ratio of the lossless manner on $conv5$ of VGGnet is only around 12x, which is not desirable for real applications.
Accordingly, lossy compression on deep features is worth for further investigation. Moreover, it has been shown that the final output result of a neural network is not sensitive to slight changes of the activations in intermediate layers \cite{han2015deep}, which provides tolerability of the information loss for the lossy compression. In addition, the dynamic range of a deep learning feature is generally much smaller than the value range of the corresponding numeric data type, which provides much room for techniques such as quantization and sampling to compress the deep learning features.
It is valuable to conduct further researches on compressing the deep learning features in a lossy way while maintaining the analysis performance.

\section{Conclusions}
\label{sec:conclusion}
We have proposed a new strategy that exploits the redundancy of intermediate deep learning features instead of visual signal or top-layer features. The advantage of this strategy lies in that the generalization capability is greatly enhanced to achieve multiple analyses tasks performed simultaneously at the cloud side, such that better trade-off can be achieved in terms of the computational load, communicational cost and generalization capability. We further conducted comprehensive lossless compression evaluations with four benchmark data compression methods on deep features of four wildly used neural networks. As the first attempt to the problem, the proposed strategy and the evaluation results in this paper provide a good reference for further studies and investigations along this vein. 



\newpage



%
\bibliographystyle{IEEEtran}
\bibliography{mybibtex}

%




\end{document}